\documentclass[journal]{IEEEtran}
\usepackage{tabularx}
\usepackage{longtable}
\usepackage{enumitem}
\usepackage{amsmath,amssymb,amsthm}
\usepackage{mathtools}
\usepackage{graphicx}
\usepackage{cite}
\usepackage{url}
\usepackage{algorithm}
\usepackage{algpseudocode}
\usepackage{booktabs}
\usepackage{tabularx}
\usepackage{booktabs}

\begin{document}

\title{Deep Q-Learning-Based Gain Scheduling for Nonlinear Quadcopter Dynamics}

\author{Hossein Rastgoftar and Muhammad J.  H. Zahed
\thanks{Authors are with the Department of Aerospace and Mechanical Engineering, University of Arizona, Tucson, AZ 85719, USA
        {\tt\small \{hrastgoftar, mjhz\}@arizona.edu}}%
}

\maketitle

\begin{abstract}
This paper presents a deep Q-network (DQN)–based gain-scheduling framework for safety-critical quadcopter trajectory tracking. Instead of directly learning control inputs, the proposed approach selects from a finite set of pre-certified stabilizing gain vectors, enabling reinforcement learning to operate within a structured and stability-preserving control architecture. By exploiting the isotropic structure of the translational dynamics, feedback gains are shared across spatial axes to reduce dimensionality while preserving performance. The learned policy adapts feedback aggressiveness in real time, applying high authority during large transients and reducing gains near convergence to limit control effort. Simulation results using a high-fidelity nonlinear quadcopter model demonstrate accurate trajectory tracking, bounded attitude excursions, smooth transition to hover after the final time, and consistent reward improvement, validating the effectiveness and robustness of the proposed learning-based gain scheduling strategy.
\end{abstract}

\begin{IEEEkeywords}
Reinforcement Learning, Deep Q-Network (DQN), Gain Scheduling, Adaptive Control, Nonlinear Control, Quadcopter Dynamics, UAV Trajectory Tracking
\end{IEEEkeywords}
\section{Introduction}
Autonomous quadcopter stabilization and hover control have been 
extensively studied using nonlinear geometric and Lyapunov-based 
formulations that provide rigorous stability guarantees about the 
hover equilibrium \cite{Lee2010,Bouabdallah2004,Mahony2012,Tayebi2006}. 
In addition, differential flatness and minimum-snap trajectory 
generation enable dynamically feasible reference tracking, with 
hover arising as a special case of constant flat outputs 
\cite{Mellinger2011}. More recently, flatness-based Snap control architectures have been developed to enlarge domains of attraction and improve tracking performance while preserving nonlinear structure \cite{ElAsslouj2023ECC}. Despite these advances, most high-performance controllers rely on fixed feedback gains that may be conservative or suboptimal across different transient regimes. Conversely, model-free reinforcement learning (RL) approaches, such as Deep Q-Networks \cite{Mnih2015}, offer adaptability but typically sacrifice interpretability and structural stability guarantees that are essential in safety-critical aerospace systems. This paper addresses the problem of integrating adaptive RL with a rigorously validated flatness-based control architecture without compromising stability structure. We propose a supervisory DQN-based gain-scheduling framework layered over the Snap controller, enabling adaptive feedback authority modulation while preserving certified nonlinear control properties. The resulting approach provides a principled bridge between RL and structured nonlinear flight control.

\subsection{Related Work}

Quadcopter trajectory tracking is often handled with nonlinear controllers that offer stability guarantees under modeling assumptions. Feedback linearization and sliding-mode designs improve robustness to uncertainty and noise \cite{lee2009feedback}. Geometric tracking on special Euclidean group SE(3) yields globally consistent attitude errors with strong tracking performance \cite{5717652}. In deployed systems, gain scheduling is common, gains are tuned for regimes or fault cases and switched online. Representative examples include gain-scheduled PID fault-tolerant control for quadrotors \cite{milhim2010gain} and parameter-dependent stability–based gain-scheduled PID for path tracking and fault tolerance on the Qball-X4 \cite{qiao2018gain}.

Compared with the above approaches, learning-based methods aim to reduce hand tuning and compensate unmodeled dynamics. A widely used architecture keeps a stabilizing feedback loop but adds a learned feedforward/reference-shaping block. Deep neural networks have been proposed as add-on modules for impromptu quadrotor trajectory tracking \cite{7989607}. Inversion-based learning extends this idea to challenging dynamics while explicitly reasoning about input–output behavior and tracking transients \cite{8279416}. Online variants combine offline training with continued adaptation during flight, incorporating expert knowledge to handle changing dynamics and operational uncertainty \cite{8794314}. Other work moves learning deeper into the controller: DATT learns adaptive components to track arbitrary (even infeasible) trajectories under large disturbances \cite{huang2023dattdeepadaptivetrajectory}. End-to-end neural controllers trained from optimal-trajectory data directly output rotor commands and address reality-gap effects by adapting to unmodeled moments \cite{FEREDE2024104588}.

RL has also been explored for low-level UAV control. GymFC evaluates policy-gradient RL for quadcopter attitude control in high-fidelity simulation and discusses robustness and reward-design challenges \cite{10.1145/3301273}. RL has enabled agile, acrobatic flight by learning sensorimotor policies in simulation with demonstration-based training \cite{kaufmann2020deep}. Earlier learning-based aerobatics include policy-gradient improvement of parametrized quadrotor multi-flips \cite{5509452}. Yet most RL flight controllers learn continuous control actions (or continuously tuned gains), and stability/constraint guarantees are typically indirect. Safe-RL formulations address constraints through constrained policy optimization \cite{pmlr-v70-achiam17a} or Lyapunov-based safety certificates \cite{chow2018lyapunov}, while deep Q-learning is a canonical value-based method for discrete action selection \cite{mnih2015human}. 

Prior gain scheduling is usually heuristic or event-triggered and lacks data-driven, state-dependent adaptation, whereas RL-based flight control often learns continuous control directly without embedding certified stability. Our approach closes this gap by using a DQN policy to schedule among a finite library of pre-certified stabilizing gain vectors, preserving stability by construction while adapting aggressiveness online. By sharing translational gains across axes (exploiting isotropy), we further reduce the learning dimension without sacrificing tracking quality. This structured action space improves exploration efficiency and simplifies verification compared with continuous-gain RL.

\subsection{Contributions}

This paper introduces a structured RL framework for nonlinear quadcopter trajectory tracking that augments, rather than replaces, certified nonlinear control architectures. The proposed method integrates a Deep Q-Network (DQN) with the Snap flatness-based controller previously introduced in \cite{ElAsslouj2023ECC, Rastgoftar2021TCNS}, whose performance was benchmarked against the widely adopted Mellinger minimum-snap controller \cite{Mellinger2011}. 
Unlike end-to-end RL approaches that directly map states to thrust and torques, the proposed framework preserves the structural properties of feedback-linearization and differential flatness while introducing adaptive gain scheduling at a supervisory level.

The main contributions of this paper are:

\begin{enumerate}

\item \textbf{Structure-Preserving RL Architecture.}  
We propose a supervisory RL formulation in which the learning agent selects stabilizing gain vectors for a validated flatness-based controller rather than generating control inputs directly. As shown in Fig. \ref{Schematic-Comp}, the proposed approach preserves stability structure and interpretability while enabling adaptive performance. The approach provides an alternative to fully model-free policies \cite{Mnih2015}, embedding learning inside a control-theoretic backbone.

\item \textbf{Safety-Conscious Hybrid Gain Scheduling with Certified Bounds.}  
The action space is restricted to pre-certified stabilizing gain configurations of the Snap controller \cite{ElAsslouj2023ECC}, and switching is regulated via dwell-time constraints. This  method ensures that adaptive scheduling does not violate the stability guarantees.

\item \textbf{Adaptive Feedback Authority without Loss of Stability Structure.}  
Simulation results demonstrate that the learned policy autonomously modulates feedback authority—aggressive gains during large transients and conservative gains near convergence—while retaining the theoretical stability properties of flatness-based control. This bridges the gap between minimum-snap trajectory tracking \cite{Mellinger2011} and adaptive RL.

\begin{figure}[t]
  \centering
  \includegraphics[width=0.82\linewidth]{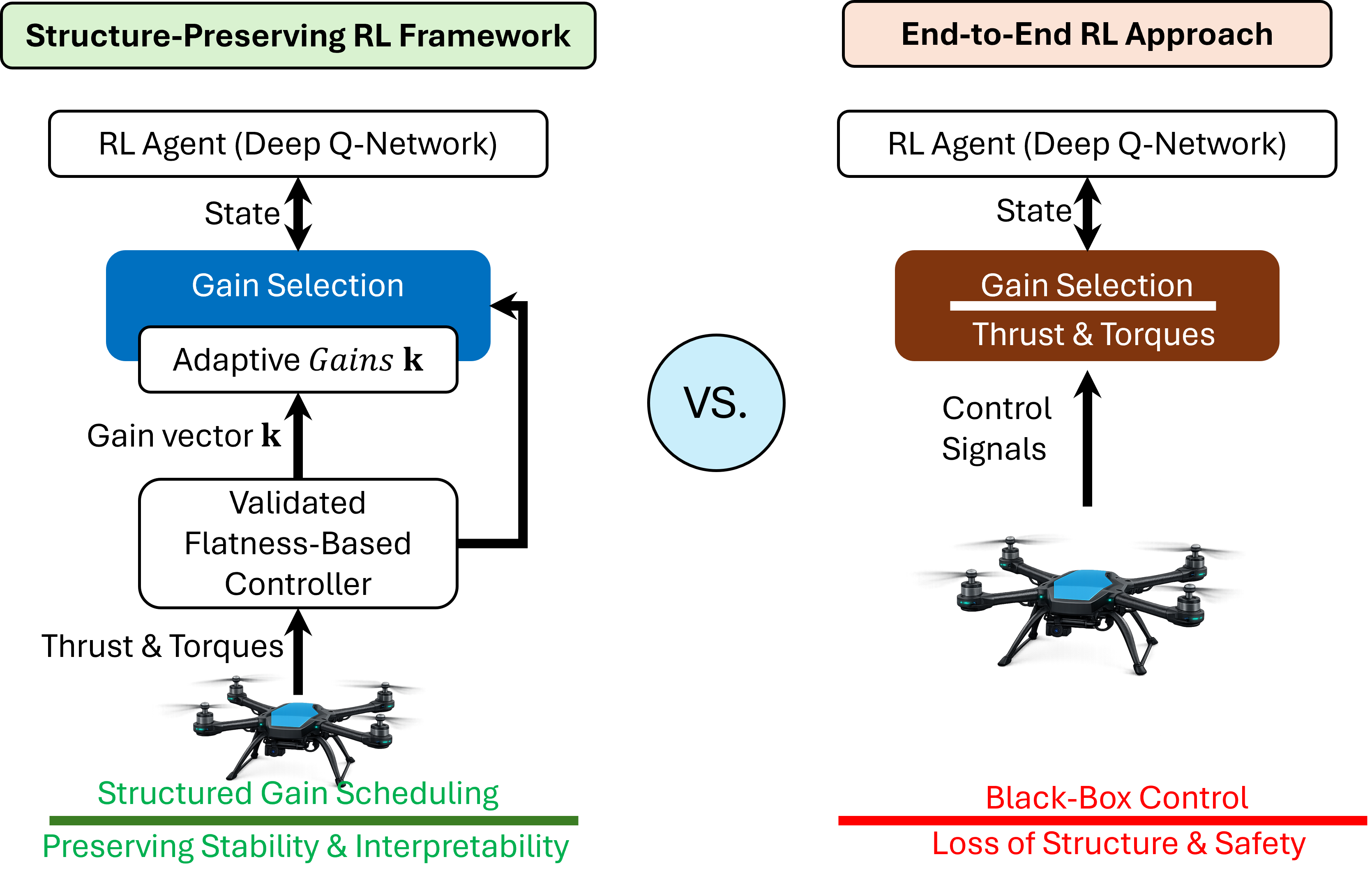}
  \caption{Comparison between structure-preserving RL and end-to-end model-free control. In the proposed architecture (left), a DQN supervises gain selection for a validated flatness-based controller, preserving stability structure and interpretability. In contrast, end-to-end RL (right) directly maps states to control inputs, potentially sacrificing structural guarantees.}
  \label{Schematic-Comp}
\end{figure}

\end{enumerate}

\subsection{Paper Outline}
The problem investigated in this paper is reviewed in Section \ref{Problem Statement}. A quadcopter control ensuring stability of a desired hover condition is presented in Section \ref{Quadcopter Model}.  Gain scheduling is presented as DQN-based RL in Section \ref{DQN-Based Gain Scheduling}. Simulation results are presented in Section \ref{Results} and followed by the concluding remarks in Section \ref{Conclusion}.

\section{Problem Statement}\label{Problem Statement}

We consider the nonlinear dynamics of a quadcopter described by
\begin{equation}\label{orginaldyn}
\begin{cases}
\dot{\mathbf{r}} = \mathbf{v}, \\[4pt]
\dot{\mathbf{v}}
=
-\,g\hat{\mathbf{e}}_3
+\dfrac{mg+T}{m}\mathbf{R}(\boldsymbol{\eta})\hat{\mathbf{e}}_3, \\[8pt]
\dot{\boldsymbol{\eta}} = \mathbf{E}(\phi,\theta)\boldsymbol{\omega}, \\[8pt]
\mathbf{I}\dot{\boldsymbol{\omega}}
+
\boldsymbol{\omega}\times(\mathbf{I}\boldsymbol{\omega})
=
\boldsymbol{\tau}, \\[8pt]
\dot T = \dot T, \\[4pt]
\ddot T = u_T,
\end{cases}
\end{equation}
where $\hat{\mathbf{e}}_3=[0~0~1]^\top$, $\mathbf{r},\mathbf{v}\in\mathbb{R}^3$ denote inertial position and velocity, 
$\boldsymbol{\eta}=(\phi,\theta,\psi)$ are the 3-2-1 Euler angles, 
$\boldsymbol{\omega}=(p,q,r)$ is the body angular velocity, 
$T$ represents thrust deviation from hover, 
$\boldsymbol{\tau}=(\tau_\phi,\tau_\theta,\tau_\psi)^\top$ is the control torque,
and $\mathbf{E}(\phi,\theta)$ is the Euler kinematic matrix.

The control inputs $\boldsymbol{\tau}$ and $u_T$ are parameterized by a gain vector
\[
\mathbf{k} = [k_1,\dots,k_{14}]^\top,
\]
where each component satisfies
\[
k_i \in [k_{i,\min}, k_{i,\max}], \quad i=1,\dots,14.
\]
The bounds are selected to prevent oscillatory behavior and ensure asymptotic convergence to a hover equilibrium. To preserve the stability properties, the gain vector is restricted to a discrete action set
\[
\mathcal{A} \subset [k_{1,\min}, k_{1,\max}] \times \cdots \times [k_{14,\min}, k_{14,\max}].
\]

Under this parametrization, the closed-loop dynamics can be expressed as
\begin{equation}\label{generalnonlinear}
\dot{\mathbf{x}} = \mathbf{f}(\mathbf{x}) + \mathbf{G}(\mathbf{x})\mathbf{k},
\end{equation}
where $\mathbf{x}\in\mathbb{R}^{14}$ denotes the full state. 

In practice, the nonlinear mappings $\mathbf{f}$ and $\mathbf{G}$ are uncertain due to modeling inaccuracies, aerodynamic disturbances, and parameter variations. Given an uncertain dynamical model and a discrete gain set $\mathcal{A}$, design a gain-selection policy
\[
\pi:\mathbf{x} \mapsto \mathbf{k} \in \mathcal{A}
\]
that guarantees safe, asymptotic convergence to a desired hover equilibrium $\mathbf{x}^\ast$, while maintaining bounded transients and avoiding oscillatory behavior.

\medskip
\noindent
This problem is formulated as a RL task, where gain scheduling is treated as a discrete-action control problem. The optimal policy is obtained by approximating the associated value function using a Deep Q-Network (DQN), thereby enabling model-free learning under uncertainty.

\section{Quacopter Control Stability}\label{Quadcopter Model}
This section reformulates the quadcopter dynamics in the form of \eqref{generalnonlinear} and develops a feedback control law that guarantees asymptotic stability of a desired hover equilibrium. First, a nonlinear state-space representation of the quadcopter dynamics is presented in Section~\ref{Nonlinear State-Space Dynamics}. Next, the stabilizing feedback control law is constructed in Section~\ref{Quadcopter Control}. Finally, Section~\ref{Control-Friendly Discrete-Time Dynamics} derives a control-oriented discrete-time model that will be used to construct the RL environment.

\subsection{Nonlinear State-Space Dynamics}\label{Nonlinear State-Space Dynamics}
Differentiating the kinematic relation
\[
\dot{\boldsymbol{\eta}}=\mathbf{E}(\phi,\theta)\boldsymbol{\omega}
\]
yields
\[
\ddot{\boldsymbol{\eta}}
=
\mathbf{E}(\phi,\theta)\dot{\boldsymbol{\omega}}
+
\dot{\mathbf{E}}(\phi,\theta)\boldsymbol{\omega}.
\]
Define the virtual control input
\[
\mathbf{u}_\eta
\triangleq
\begin{bmatrix}
u_\phi\\
u_\theta\\
u_\psi
\end{bmatrix}
:=\ddot{\boldsymbol{\eta}}.
\]
Solving for $\dot{\boldsymbol{\omega}}$ gives
\begin{equation}
    \dot{\boldsymbol{\omega}}
=
\mathbf{E}^{-1}(\phi,\theta)
\Bigl(
\mathbf{u}_\eta
-
\dot{\mathbf{E}}(\phi,\theta)\boldsymbol{\omega}
\Bigr).
\end{equation}
Substituting into the rigid-body rotational dynamics yields the required control torque
\[
\boldsymbol{\tau}
=
\mathbf{I}\mathbf{E}^{-1}(\phi,\theta)\mathbf{u}_\eta
-
\mathbf{I}\mathbf{E}^{-1}(\phi,\theta)\dot{\mathbf{E}}(\phi,\theta)\boldsymbol{\omega}
+
\boldsymbol{\omega}\times(\mathbf{I}\boldsymbol{\omega}).
\]
By defining
\[
\begin{split}
    \mathbf{x}=&\begin{bmatrix}
\mathbf{r}^\top & \mathbf{v}^\top & \boldsymbol{\eta}^\top & \dot{\boldsymbol{\eta}}^\top & T & \dot{T}
\end{bmatrix}^\top\in\mathbb{R}^{14},\\
\mathbf{u}=&\begin{bmatrix}
u_T & \mathbf{u}_\eta^\top
\end{bmatrix}^\top\in\mathbb{R}^{4},
\end{split}
\]
and using $\boldsymbol{\omega}=\mathbf{E}^{-1}(\phi,\theta)\dot{\boldsymbol{\eta}}$, the quadcopter dynamics admit the control-affine form
\begin{equation}\label{dyn0}    \dot{\mathbf{x}}=\mathbf{f}_0\!\left(\mathbf{x}\right)+\mathbf{G}_0\!\left(\mathbf{x}\right)\mathbf{u}.
\end{equation}
where 
\begin{equation}
\mathbf{f}_0\!\left(\mathbf{x}\right)=
\begin{bmatrix}
\mathbf{v}_I \\[4pt]
-\,g\hat{\mathbf{e}}_3+\dfrac{mg+T}{m}\mathbf{R}(\boldsymbol{\eta})\hat{\mathbf{e}}_3\\[8pt]
\dot{\boldsymbol{\eta}} \\[4pt]
\mathbf{0}_{3} \\[6pt]
\dot T \\[4pt]
0
\end{bmatrix},
\end{equation}

\begin{equation}
\mathbf{G}_0\!\left(\mathbf{x}\right)=
\begin{bmatrix}
\mathbf{0}_{9\times 1}&\mathbf{0}_{9\times 3}\\
\mathbf{0}_{3\times 1} & \mathbf{I}_3\\
0 & \mathbf{0}_{1\times 3}\\
1 & \mathbf{0}_{1\times 3}
\end{bmatrix}.
\end{equation}

\subsection{Quadcopter Control}\label{Quadcopter Control}

We use the snap control framework developed in \cite{el2023quadcopter} to achieve a desired hover equilibrium of the form
\[
\mathbf{x}^*=
\begin{bmatrix}
\mathbf{r}_I^*\\
\mathbf{0}_{11\times 1}
\end{bmatrix},
\]
where $\mathbf{r}_I^*$ denotes the desired hover position. To ensure a smooth transition from an arbitrary initial position $\mathbf{r}_0$ to $\mathbf{r}_I^*$, we define the desired trajectory
\begin{equation}\label{rd}
\mathbf{r}_d(t)=
\begin{cases}
(1-\beta(t))\mathbf{r}_0+\beta(t)\mathbf{r}_I^*,
& t\in[0,T_f],\\[4pt]
\mathbf{r}_I^*,
& t\ge T_f,
\end{cases}
\end{equation}
where $\beta(t)$ is a quintic polynomial satisfying
\[
\beta(0)=0,\quad
\beta(T_f)=1,\quad
\dot\beta(T_f)=0,
\]
\[
\ddot\beta(T_f)=0,\quad
\dddot\beta(T_f)=0,\quad
\ddddot\beta(T_f)=0.
\]
These conditions enforce $C^4$ continuity at $t=T_f$ under the post-$T_f$ constant hold.

Defining $\mathbf{e}_r=\mathbf{r}-\mathbf{r}_d$, $\mathbf{e}_v=\mathbf{v}-\dot{\mathbf{r}}_d$, $\mathbf{e}_a=\mathbf{a}-\ddot{\mathbf{r}}_d$, $\mathbf{e}_j=\mathbf{j}-\dddot{\mathbf{r}}_d$,    the external (hover-error) state vector is given by
\[
\mathbf{z}(\mathbf{x})
\triangleq
\begin{bmatrix}
\mathbf{e}_r^\top &
\mathbf{e}_v^\top &
\mathbf{e}_a^\top &
\mathbf{e}_j^\top &
\psi &
\dot\psi
\end{bmatrix}^\top
\in\mathbb{R}^{14}.
\]

Let the external input be
\[
\mathbf{s}
\triangleq
\begin{bmatrix}
\mathbf{s}_r\\
s_\psi
\end{bmatrix}
=
\begin{bmatrix}
\dot{\mathbf{j}}\\
\ddot\psi
\end{bmatrix}
\in\mathbb{R}^{4},
\]
where $\mathbf{s}_r$ is the translational snap input and $s_\psi$ is the yaw angular acceleration. Then, the external dynamics take the form
\begin{equation}
\dot{\mathbf{z}}
=
\mathbf{A}_{\mathrm{EXT}}\mathbf{z}
+
\mathbf{B}_{\mathrm{EXT}}\mathbf{s}
+
\mathbf{E}_{\mathrm{EXT}}\mathbf{r}_d^{(4)}(t),
\end{equation}
where $\mathbf{r}_d^{(4)}(t)$ denotes the desired snap and

\[
\mathbf{A}_{\mathrm{EXT}}=
\begin{bmatrix}
\mathbf{0}_{3} & \mathbf{I}_{3} & \mathbf{0}_{3} & \mathbf{0}_{3} & \mathbf{0}_{3\times1} & \mathbf{0}_{3\times1}\\
\mathbf{0}_{3} & \mathbf{0}_{3} & \mathbf{I}_{3} & \mathbf{0}_{3} & \mathbf{0}_{3\times1} & \mathbf{0}_{3\times1}\\
\mathbf{0}_{3} & \mathbf{0}_{3} & \mathbf{0}_{3} & \mathbf{I}_{3} & \mathbf{0}_{3\times1} & \mathbf{0}_{3\times1}\\
\mathbf{0}_{3} & \mathbf{0}_{3} & \mathbf{0}_{3} & \mathbf{0}_{3} & \mathbf{0}_{3\times1} & \mathbf{0}_{3\times1}\\
\mathbf{0}_{1\times3} & \mathbf{0}_{1\times3} & \mathbf{0}_{1\times3} & \mathbf{0}_{1\times3} & 0 & 1\\
\mathbf{0}_{1\times3} & \mathbf{0}_{1\times3} & \mathbf{0}_{1\times3} & \mathbf{0}_{1\times3} & 0 & 0
\end{bmatrix},
\]

\[
\mathbf{B}_{\mathrm{EXT}}=
\begin{bmatrix}
\mathbf{0}_{3\times3} & \mathbf{0}_{3\times1}\\
\mathbf{0}_{3\times3} & \mathbf{0}_{3\times1}\\
\mathbf{0}_{3\times3} & \mathbf{0}_{3\times1}\\
\mathbf{I}_{3}        & \mathbf{0}_{3\times1}\\
\mathbf{0}_{1\times3} & 0\\
\mathbf{0}_{1\times3} & 1
\end{bmatrix},
\qquad 
\mathbf{E}_{\mathrm{EXT}}=
\begin{bmatrix}
\mathbf{0}_{3\times3}\\
\mathbf{0}_{3\times3}\\
\mathbf{0}_{3\times3}\\
-\mathbf{I}_{3}\\
\mathbf{0}_{1\times3}\\
\mathbf{0}_{1\times3}
\end{bmatrix}.
\]

The translational subsystem thus forms a fourth-order chain of integrators driven by snap, while the yaw dynamics form a second-order subsystem driven by $s_\psi$. 
We choose

\begin{equation}
\begin{split}
    \mathbf{s}=&-k_{13}\dot{\psi}-k_{14}\psi
+\mathbf{K}_j\left(\dddot{\mathbf{r}}_d-\mathbf{j}\right)
+\mathbf{K}_a\left(\ddot{\mathbf{r}}_d-\mathbf{a}\right)\\
+&\mathbf{K}_v\left(\dot{\mathbf{r}}_d-\mathbf{v}\right)
+\mathbf{K}_p\left({\mathbf{r}}_d-\mathbf{r}\right)
\end{split}    
\end{equation}
with positive definite diagonal gain matrices
\[
\begin{split}
    \mathbf{K}_j=&\mathrm{diag}(k_1,k_2,k_3),\\
\mathbf{K}_a=&\mathrm{diag}(k_4,k_5,k_6),\\
\mathbf{K}_v=&\mathrm{diag}(k_7,k_{8},k_{9}),\\
\mathbf{K}_p=&\mathrm{diag}(k_{10},k_{11},k_{12}),
\end{split}
\]
and yaw gains $k_{13},k_{14}>0$ , where
\begin{equation}
    k_i=\left[k_{i,min},k_{i,max}\right],\qquad i=1,\cdots,14.
\end{equation}
Note that the bounds $k_{i,\min}$ and $k_{i,\max}$ listed in Table~\ref{tab:gain_bounds} are selected such that the resulting closed-loop external dynamics matrix has distinct eigenvalues strictly in the open left half-plane, thereby guaranteeing asymptotic stability.

\noindent\textbf{Relation between external input $\mathbf{s}$ and quadcopter virtual input $\mathbf{u}$.}
Define
\begin{equation}
\hat{\mathbf{k}}_b=\mathbf{R}(\boldsymbol{\eta})\hat{\mathbf{e}}_3
\end{equation}
as the unit normal vector to the quadcopter plane, aligned with the thrust direction, $\hat{\mathbf{e}}_3$ denotes the third canonical base vector in $\mathbb{R}^3$.
By differentiating the translational dynamics twice and using the kinematic relations
\[
\dot{\boldsymbol{\eta}}=\mathbf{E}(\phi,\theta)\boldsymbol{\omega},
\qquad
\boldsymbol{\omega}=\mathbf{E}^{-1}(\phi,\theta)\dot{\boldsymbol{\eta}},
\]
the external input $\mathbf{s}$ can be written as an affine function of the virtual input $\mathbf{u}$:
\begin{equation}
\mathbf{s}
=
\mathbf{M}(\mathbf{x})\,\mathbf{u}
+
\mathbf{n}(\mathbf{x}),
\end{equation}
where
\begin{equation}\label{su}
\mathbf{M}(\mathbf{x})
=
\begin{bmatrix}
\mathbf{M}'(\mathbf{x})\\
0\;\;0\;\;0\;\;1
\end{bmatrix}
\in\mathbb{R}^{4\times 4},
\qquad
\mathbf{n}(\mathbf{x})
=
\begin{bmatrix}
\mathbf{n}'(\mathbf{x})\\
0
\end{bmatrix}
\in\mathbb{R}^{4}.
\end{equation}

The $3\times 4$ matrix $\mathbf{M}'(\mathbf{x})$ is given explicitly by
\begin{equation}
\resizebox{0.99\hsize}{!}{%
$
\mathbf{M}'(\mathbf{x})
=
\begin{bmatrix}
\dfrac{1}{m}\hat{\mathbf{k}}_b &
f\left[\mathbf{E}^{-1}\hat{\mathbf e}_1\right]_{\times}\hat{\mathbf{k}}_b &
f\left[\mathbf{E}^{-1}\hat{\mathbf e}_2\right]_{\times}\hat{\mathbf{k}}_b &
f\left[\mathbf{E}^{-1}\hat{\mathbf e}_3\right]_{\times}\hat{\mathbf{k}}_b
\end{bmatrix},
$
}
\end{equation}
and the drift term $\mathbf{n}'(\mathbf{x})$ is
\begin{equation}
\mathbf{n}'(\mathbf{x})
=
-f\left[\mathbf{E}^{-1}\dot{\mathbf E}\boldsymbol{\omega}\right]_{\times}\hat{\mathbf{k}}_b
+
2\frac{\dot T}{m}\left[\boldsymbol{\omega}\right]_{\times}\hat{\mathbf{k}}_b
+
f\left[\boldsymbol{\omega}\right]_{\times}
\left[\boldsymbol{\omega}\right]_{\times}\hat{\mathbf{k}}_b,
\end{equation}
where 
\begin{equation}
    f= \dfrac{mg+T}{m}.
\end{equation}
Here $[\cdot]_{\times}$ denotes the skew--symmetric matrix associated with the cross product,
and $\hat{\mathbf e}_i$ denotes the $i$th canonical basis vector in $\mathbb{R}^3$, i.e.,
\[
\hat{\mathbf e}_1=[1\;0\;0]^\top,\qquad
\hat{\mathbf e}_2=[0\;1\;0]^\top,\qquad
\hat{\mathbf e}_3=[0\;0\;1]^\top.
\]

\subsection{Control-Friendly Discrete-Time Dynamics}
\label{Control-Friendly Discrete-Time Dynamics}

The external input can be written compactly as
\begin{equation}
\mathbf{s}=\mathbf{H}(\mathbf{x})\mathbf{k},
\end{equation}
where $\mathbf{H}:\mathbb{R}^{14}\to\mathbb{R}^{4\times14}$. From \eqref{su}, the virtual control input is
\begin{equation}
\mathbf{u}
=
\mathbf{M}^{-1}\!\left(\mathbf{H}\mathbf{k}-\mathbf{n}\right),
\end{equation}
which, when substituted into \eqref{dyn0}, yields the control-affine form
\begin{equation}
\dot{\mathbf{x}}
=
\mathbf{f}(\mathbf{x})
+
\mathbf{G}(\mathbf{x})\mathbf{k},
\label{eq:control_affine}
\end{equation}
with
\begin{equation}
\mathbf{f}
=
\mathbf{f}_0
-
\mathbf{G}_0\mathbf{M}^{-1}\mathbf{n},
\qquad
\mathbf{G}
=
\mathbf{G}_0\mathbf{M}^{-1}\mathbf{H}.
\end{equation}
Under zero-order hold with sampling time $\Delta t$, the discrete-time dynamics are
\begin{equation}
\mathbf{x}_{t+1}
=
\Phi(\mathbf{x}_t,\mathbf{k}_t),
\end{equation}
where $\Phi$ is computed numerically via a fourth-order Runge--Kutta scheme applied to \eqref{eq:control_affine}.

\section{DQN-Based Gain Scheduling}\label{DQN-Based Gain Scheduling}

We cast the gain scheduling problem as a Markov decision process (MDP) over the discrete-time system
\begin{equation}\label{closed-loop}
\mathbf{x}_{t+1}
=
\Phi(\mathbf{x}_t,a_t),
\quad
a_t \in \{1,\dots,N\},
\end{equation}
where each action selects a gain vector from the finite stabilizing set
\begin{equation}
\mathcal{A}
=
\{\mathbf{k}^{(1)},\dots,\mathbf{k}^{(N)}\}
\subset
\mathcal{K}_1 \times \cdots \times \mathcal{K}_{14},
\end{equation}
where $\mathcal{K}_i\subset \left[k_{i,min},k_{i,max}\right]$ is a discrete specifying admissible values of $k_i$.
Each $\mathbf{k}\in\mathcal{A}$ renders the nominal error dynamics asymptotically stable. To prevent destabilizing rapid switching between stabilizing gain vectors,
a discrete dwell-time constraint is enforced. Specifically, whenever a new
gain vector $\mathbf{k}_t$ is selected, it must be held constant for at least
$N_d$ sampling intervals, i.e.,
\begin{equation}\label{dwell}
\mathbf{k}_{t+i} = \mathbf{k}_t,
\quad i=1,\dots,N_d.
\end{equation}
This constraint prevents high-frequency switching and preserves
stability of the resulting switched closed-loop system.
\subsection{MDP Formulation}

The gain scheduling problem is formulated as an MDP 
defined by the tuple
\[
(\mathcal{S},\mathcal{A},\mathcal{P},\mathcal{R}),
\]
where the state space $\mathcal{S}$, action space $\mathcal{A}$,
transition kernel $\mathcal{P}$, and reward function $\mathcal{R}$
are defined as follows.

\noindent\textbf{State space $\mathcal{S}$:}
The state (observation) at time $t$ is
\begin{equation}
\mathbf{o}_t
=
\begin{bmatrix}
\mathbf{x}_t \\
\phi_t
\end{bmatrix}
\in \mathcal{S},
\end{equation}
where $\mathbf{x}_t\in\mathbb{R}^{14}$ is the quadcopter state and
\[
\phi_t = \min(t/T_f,1)
\]
is a phase variable encoding the progression of the reference trajectory.
The inclusion of $\phi_t$ preserves the Markov property under time-varying
reference tracking.

\medskip

\noindent\textbf{Action space $\mathcal{A}$:}
The action is a discrete index
\begin{equation}
a_t \in \{1,\dots,N\},
\end{equation}
selecting the gain vector $\mathbf{k}^{(a_t)}\in\mathcal{A}$, where
\[
\mathcal{A}
=
\{\mathbf{k}^{(1)},\dots,\mathbf{k}^{(N)}\}
\subset
\mathcal{K}_1 \times \cdots \times \mathcal{K}_{14}.
\]
Each $\mathbf{k}\in\mathcal{A}$ renders the nominal error dynamics
asymptotically stable. A dwell-time constraint, given by Eq. \eqref{dwell},
is enforced to prevent destabilizing rapid switching.

\medskip

\noindent\textbf{Transition kernel $\mathcal{P}$:}
The state transition probability is induced by the discrete-time
closed-loop dynamics \eqref{closed-loop}
where $\Phi$ is obtained by numerically integrating the
control-affine system under zero-order hold.
Modeling uncertainty and unmodeled aerodynamic effects
render the transition kernel $\mathcal{P}$ unknown.

\medskip

\noindent\textbf{Reward function $\mathcal{R}$:}
The stage reward penalizes tracking errors, attitude deviation,
angular velocity, control effort, and gain switching:
\begin{equation}
\begin{aligned}
r_t = - \Big(
& w_r \|\mathbf{e}_r\|^2
+ w_v \|\mathbf{e}_v\|^2
+ w_\eta \|\boldsymbol{\eta}\|^2
+ w_\omega \|\boldsymbol{\omega}\|^2
\Big) \\
& - w_u \|\mathbf{u}_t\|^2
- w_s \mathbb{I}[a_t \neq a_{t-1}].
\end{aligned}
\end{equation}
where $w_r$, $w_v$, $w_\eta$, $w_\omega$, $w_u$, and $w_s$ are positive scaling weights.
\subsection{Deep Q-Network Approximation}

Let $Q^\pi(\mathbf{o},a)$ denote the action-value function.
The optimal Q-function satisfies the Bellman equation
\begin{equation}
Q^\star(\mathbf{o}_t,a_t)
=
\mathbb{E}
\left[
r_t
+
\gamma
\max_{a'}
Q^\star(\mathbf{o}_{t+1},a')
\right].
\end{equation}
To approximate $Q^\star$, we employ a Deep Q-Network (DQN)
\[
Q_\theta : \mathbb{R}^{15} \rightarrow \mathbb{R}^{|\mathcal{A}|}.
\]

The network consists of two fully connected hidden layers with ReLU
activation functions, followed by a linear output layer.
This architecture provides sufficient expressive capacity
to approximate the nonlinear action-value function
while maintaining computational efficiency.

\noindent\textbf{Structured Gain Parameterization:}
A robust discrete action set is constructed by scaling nominal pole locations over a small grid and including multiple yaw pole pairs. The resulting gain table defines a family of stabilizing controllers; the bounds $k_{i,\min}$ and $k_{i,\max}$ in Table~\ref{tab:gain_bounds} are the componentwise minimum and maximum over this table. By construction, each admissible action corresponds to a controller that stabilizes the nominal error dynamics when held constant, while providing sufficient variability for RL-based performance adaptation under uncertainty.

To reduce the dimensionality of the gain-selection problem, we exploit the symmetric chain-of-integrators structure of the translational error dynamics and enforce equality of gains across the $x$, $y$, and $z$ channels at each derivative level:
\begin{equation}
k_{(3l-1)+1}=k_{(3l-1)+2}=k_{3l}, \qquad l=1,2,3,4.
\end{equation}
These correspond to jerk-, acceleration-, velocity-, and position-level feedback gains, respectively. The yaw gains $(k_1,k_2)$ remain independent due to the distinct second-order yaw dynamics and are fixed.

For each derivative level, five admissible gain values are selected, yielding a discrete action set of cardinality
\[
|\mathcal{A}| = 5^4 = 625.
\]
Accordingly, the DQN outputs a vector of 625 Q-values, where the $a$-th entry represents the value of selecting the corresponding gain vector.


\noindent\textbf{Loss Function:} The network parameters $\theta$ are updated by minimizing the
temporal-difference loss
\begin{equation}
\mathcal{L}(\theta)
=
\mathbb{E}
\Big[
\big(
Q_\theta(\mathbf{o}_t,a_t)
-
y_t
\big)^2
\Big],
\end{equation}
where
\begin{equation}
y_t
=
r_t
+
\gamma
\max_{a'}
Q_{\bar{\theta}}(\mathbf{o}_{t+1},a').
\end{equation}
Here $Q_{\bar{\theta}}$ denotes a periodically updated target network.
Experience replay is employed to decorrelate samples and improve
training stability.

\subsection{Learned Gain Scheduling Policy}

The resulting gain-scheduling policy is
\begin{equation}
\pi(\mathbf{o}_t)
=
\arg\max_{a}
Q_\theta(\mathbf{o}_t,a).
\end{equation}
Because each action corresponds to a stabilizing gain vector and switching is regulated via dwell-time, the learned scheduler preserves nominal stability while adapting gains online to compensate for model uncertainty and transient dynamics.
\begin{table*}[t]
\centering
\caption{Gain bounds derived from the stabilizing gain table used by the training environment.}
\label{tab:gain_bounds}
\renewcommand{\arraystretch}{1.15}
\setlength{\tabcolsep}{4pt}
\begin{tabular}{lcccccccccccccc}
\toprule
 & $k_1$ & $k_2$ & $k_3$ & $k_4$ & $k_5$ & $k_6$ & $k_7$ & $k_8$ & $k_9$ & $k_{10}$ & $k_{11}$ & $k_{12}$ & $k_{13}$ & $k_{14}$ \\
\midrule
$k_{i,\min}$ & 9.8304 & 24.1920 & 49.1520 & 25.6000 & 47.6160 & 78.8480 & 22.4000 & 32.9600 & 45.4400 & 8.0000 & 9.6000 & 11.2000 & 12.0000 & 8.0000 \\
$k_{i,\max}$ & 49.7664 & 122.4720 & 248.8320 & 86.4000 & 160.7040 & 266.1120 & 50.4000 & 74.1600 & 102.2400 & 12.0000 & 14.4000 & 16.8000 & 32.0000 & 12.0000 \\
\bottomrule
\end{tabular}
\end{table*}

\section{Simulation Results}
\label{Results}
We evaluate the proposed DQN-based gain-scheduling controller in a high-fidelity
black-box simulation of a quadcopter with Euler ZYX attitude and thrust/torque
actuation. The physical state is
$
\mathbf{x} =
[\mathbf{r}^\top,\mathbf{v}^\top,\boldsymbol{\eta}^\top,\boldsymbol{\omega}^\top,T_{\mathrm{dev}},\dot T]^\top
\in \mathbb{R}^{14},
$
where $\mathbf{r},\mathbf{v}\in\mathbb{R}^3$ are inertial position/velocity,
$\boldsymbol{\eta}=[\phi,\theta,\psi]^\top$ are Euler angles,
$\boldsymbol{\omega}\in\mathbb{R}^3$ are body rates,
and $T_{\mathrm{dev}}$ and $\dot T$ model thrust deviation dynamics.
The simulated vehicle parameters are fixed to
mass $m=1.5~\mathrm{kg}$, gravitational acceleration $g=9.81~\mathrm{m/s^2}$,
and inertia matrix $I=\mathrm{diag}(0.02,\,0.02,\,0.04)~\mathrm{kg\,m^2}$.
The integration step is $\Delta t = 0.01~\mathrm{s}$, with episodes of length
$10~\mathrm{s}$. The reference trajectory is generated by a smooth quintic
time-scaling over $t\in[0,T_f]$ with $T_f=5~\mathrm{s}$; for $t>T_f$ the desired
position is held constant at $\mathbf{r}_d(T_f)$ and the desired velocity,
acceleration, jerk, and snap are set to zero.

The DQN observation concatenates the 14-dimensional physical state and a scalar
phase variable $\min(t/T_f,1)$, yielding a 15-dimensional input.
At each step the agent selects a discrete action corresponding to one
pre-computed stabilizing gain vector
$
\mathbf{k}=[k_1,\dots,k_{14}]^\top
$
(from a finite table). To reduce the action space and exploit the assumed
isotropy of the translational channels, gains are shared across axes at each
derivative level (position/velocity/acceleration/jerk), while yaw gains remain
independent due to the distinct second-order yaw dynamics. A dwell-time
constraint enforces that the selected action is held for a fixed number of
steps to prevent destabilizing rapid switching.

\begin{figure}[t]
  \centering
  \includegraphics[width=0.82\linewidth]{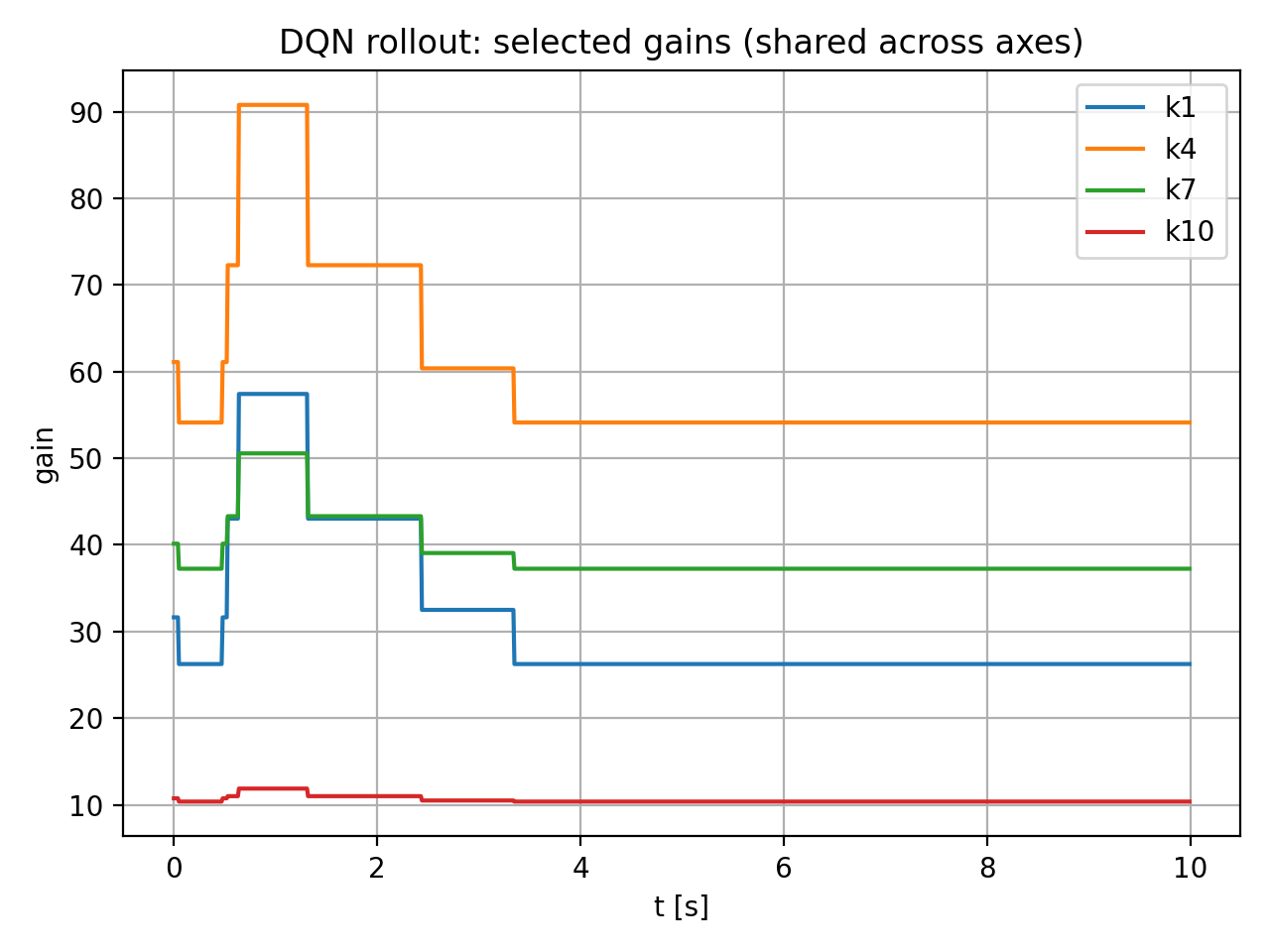}
  \caption{DQN rollout: selected translational feedback gains shared across
  axes (representative gains shown). The policy increases gains during the
  initial transient and reduces them as tracking errors diminish.}
  \label{fig:dqn_gains}
\end{figure}

Figure~\ref{fig:dqn_gains} shows the gains selected by the trained DQN during a
representative rollout (only the shared translational gains are plotted). The
policy initially chooses more aggressive feedback to quickly suppress large
initial tracking errors, then transitions to a lower-gain regime as the state
approaches the reference and the episode enters the post-$T_f$ hold phase.

\begin{figure}[t]
  \centering
  \includegraphics[width=0.95\linewidth]{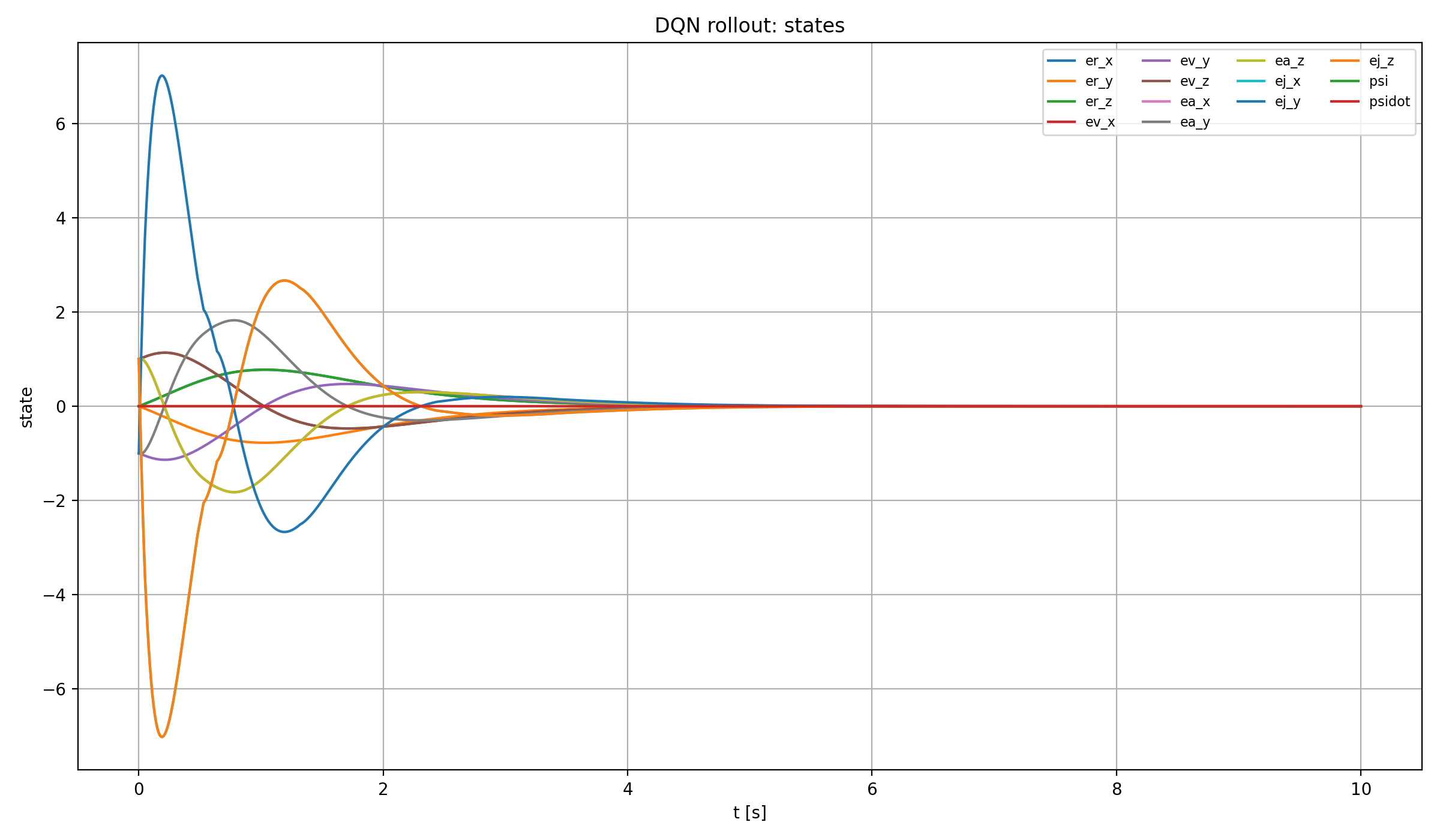}
  \caption{DQN rollout: external error states. Translational error components
  (position/velocity/acceleration/jerk) converge toward the origin while the yaw
  channel remains regulated, demonstrating stable closed-loop behavior under
  learned gain scheduling.}
  \label{fig:dqn_states}
\end{figure}

The resulting external error-state evolution is reported in
Figure~\ref{fig:dqn_states}, where the position-, velocity-, acceleration-, and
jerk-level error components in $(x,y,z)$ decay rapidly toward the origin, and
the yaw channel remains well regulated. This behavior is consistent with the
intended role of gain scheduling: applying high authority when far from the
target and relaxing control effort near convergence.

\begin{figure}[t]
  \centering
  \includegraphics[width=0.92\linewidth]{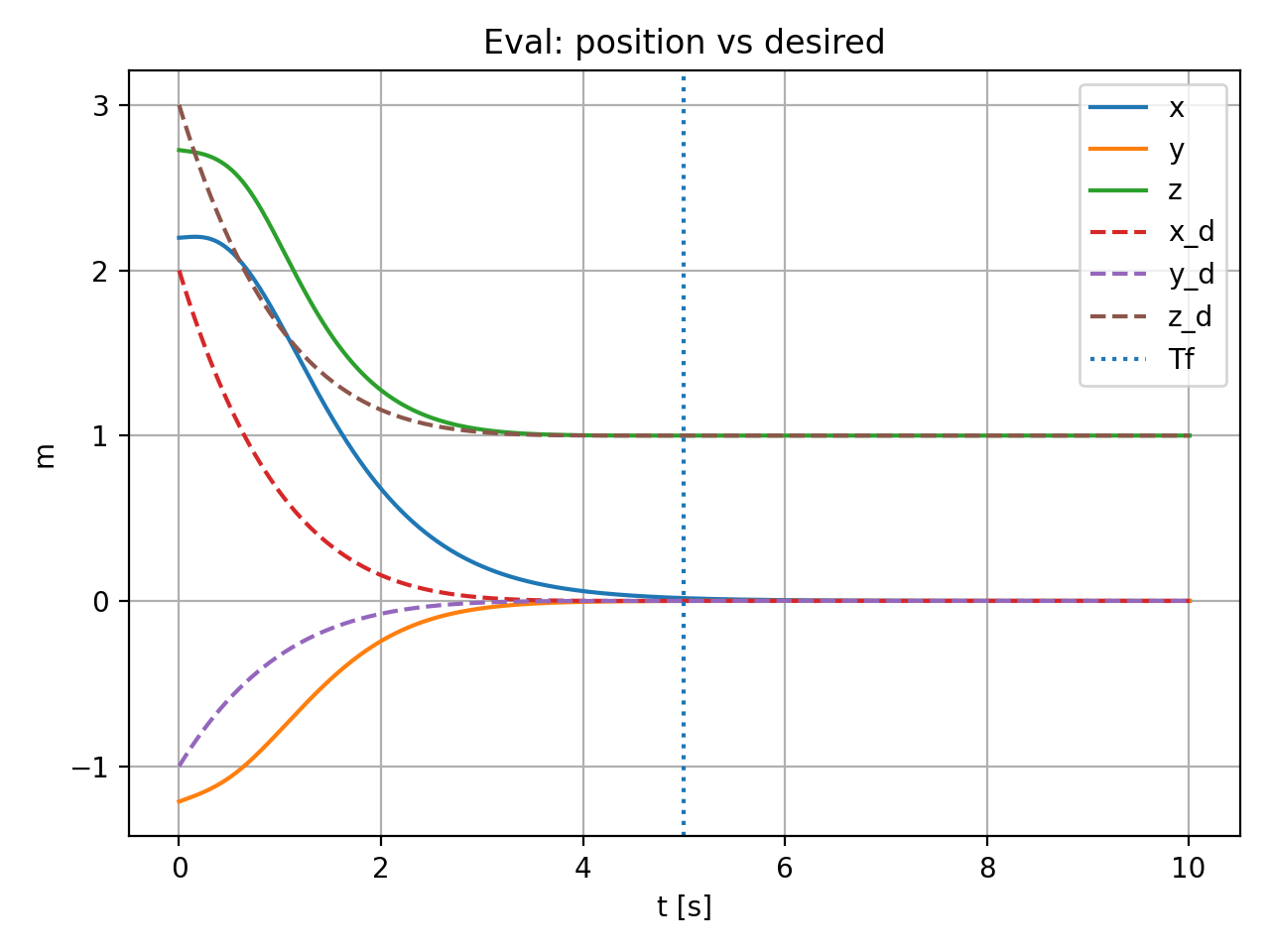}
  \caption{Physical evaluation: inertial position versus desired position.
  The dotted line marks $T_f$; for $t>T_f$ the reference is held at
  $\mathbf{r}_d(T_f)$ and the quadcopter settles to hover.}
  \label{fig:pos_vs_des}
\end{figure}

\begin{figure}[t]
  \centering
  \includegraphics[width=0.78\linewidth]{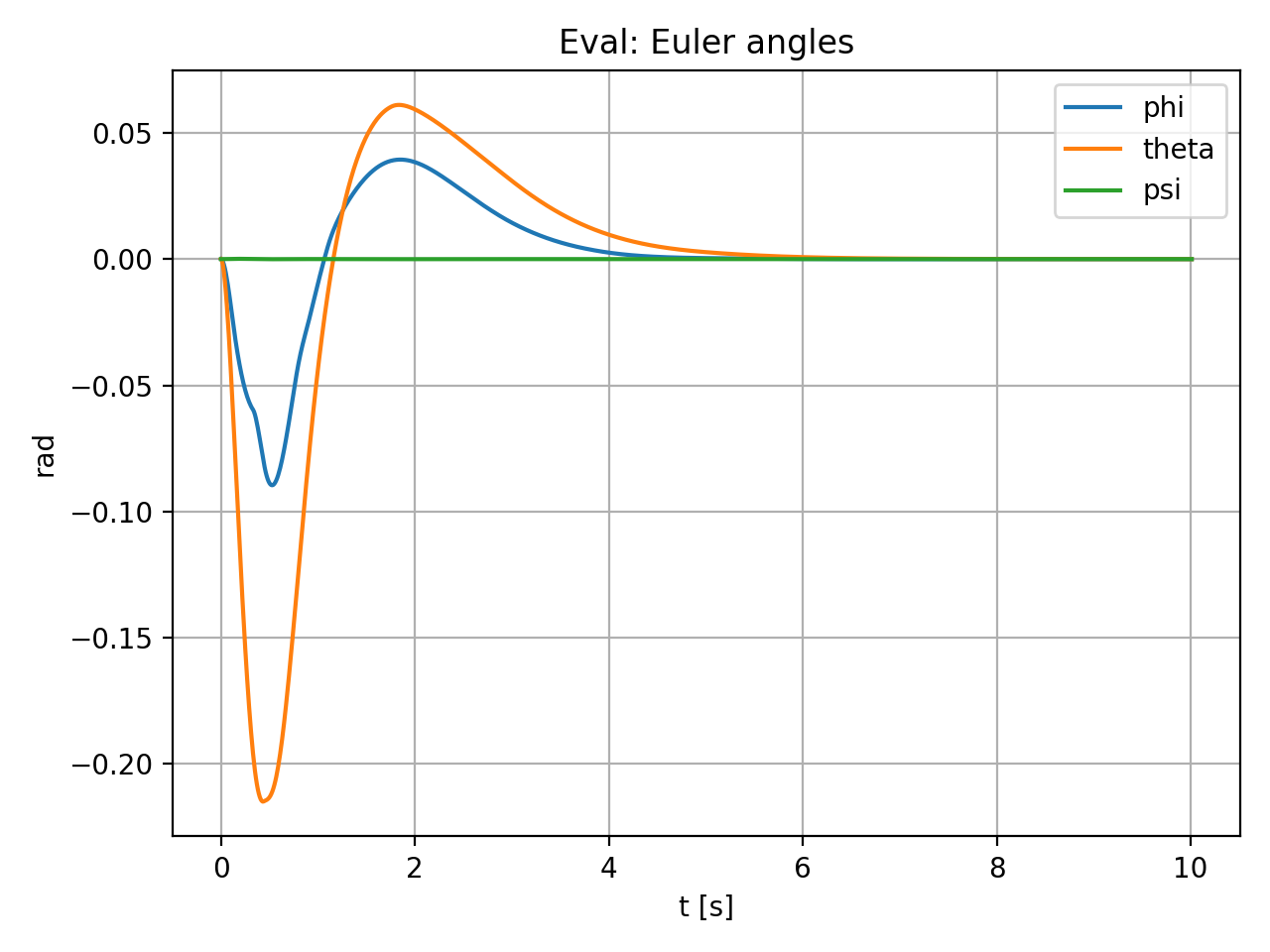}
  \caption{Physical evaluation: Euler angles $(\phi,\theta,\psi)$. Attitude
  excursions remain small and decay to near zero as tracking converges.}
  \label{fig:euler}
\end{figure}

To connect the learned scheduling to physical motion, Figure~\ref{fig:pos_vs_des}
compares the inertial position $\mathbf{r}(t)$ to the desired trajectory
$\mathbf{r}_d(t)$, with a vertical marker at $T_f$. The quadcopter tracks the
quintic segment and smoothly settles to the final hover position after $T_f$.
Figure~\ref{fig:euler} reports the Euler angles, confirming that attitude
excursions remain small during the transient and decay to near-zero steady-state.

\begin{figure}[t]
  \centering
  \includegraphics[width=0.82\linewidth]{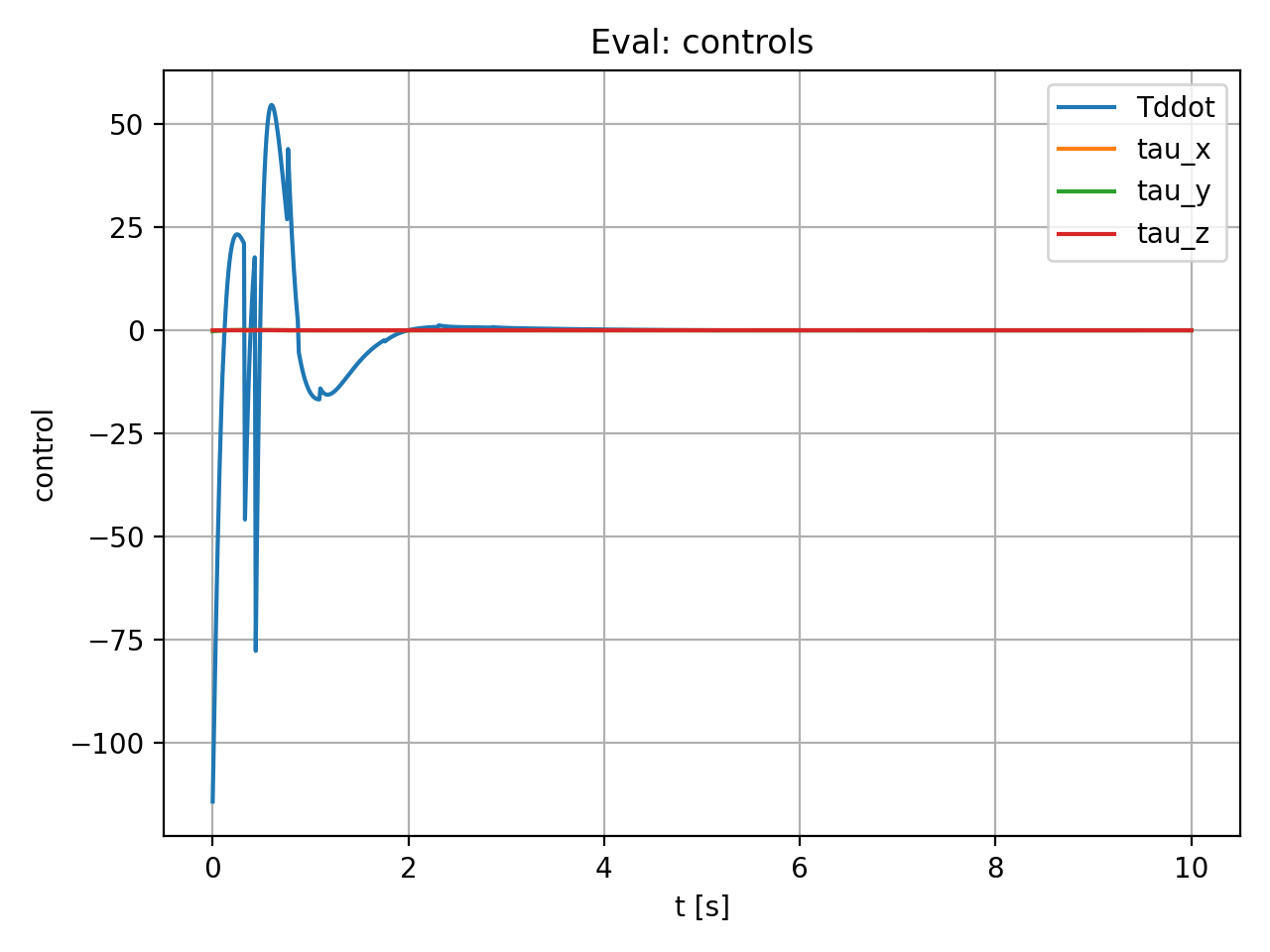}
  \caption{Physical evaluation: control inputs. The thrust second-derivative
  command $\ddot T$ and body torques $\boldsymbol{\tau}$ are largest during the
  initial transient and decrease as the state approaches the reference.}
  \label{fig:controls}
\end{figure}

\begin{figure}[t]
  \centering
  \includegraphics[width=0.82\linewidth]{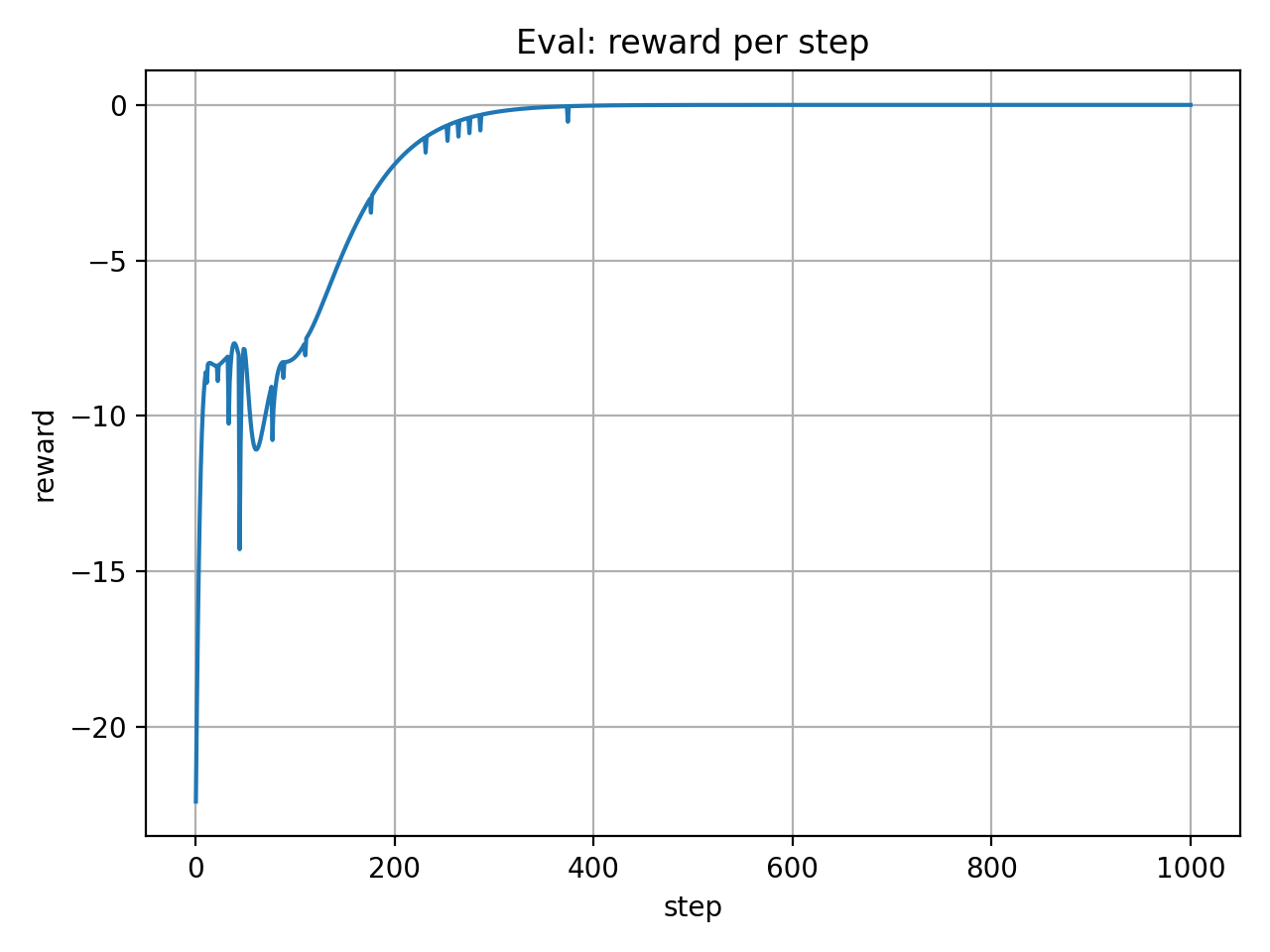}
  \caption{Physical evaluation: reward per step. The reward improves toward
  zero as tracking errors and control effort diminish over the episode.}
  \label{fig:reward}
\end{figure}

Figure~\ref{fig:controls} depicts the physical control signals used in the
simulation: the thrust second derivative command $\ddot T$ and body torques
$\boldsymbol{\tau}=[\tau_x,\tau_y,\tau_z]^\top$. As expected, the largest control
magnitudes occur early in the episode when the tracking errors are largest, and
then diminish as the trajectory is reached. Finally, the per-step reward in
Figure~\ref{fig:reward} improves from large negative values during the initial
correction to values near zero as the state converges and the policy reduces
unnecessary control effort, indicating successful stabilization and tracking
under the learned schedule.

\section{Conclusion}\label{Conclusion}
A DQN-based gain-selection strategy for quadcopter trajectory tracking has been developed and evaluated in nonlinear simulation. By restricting the action space to stabilizing gain vectors and enforcing shared translational gains, the approach achieves dimensionality reduction while maintaining structural consistency with the underlying dynamics. The learned policy demonstrates adaptive behavior, increasing feedback authority during large tracking errors and reducing gains as the system converges, thereby balancing performance and control effort. Simulation results confirm stable closed-loop operation, accurate reference tracking, bounded control inputs, and smooth hover stabilization after the terminal time. These results indicate that RL can be integrated with structured feedback architectures to achieve adaptive, safety-conscious performance without sacrificing interpretability or stability guarantees.

\bibliographystyle{IEEEtran}
\bibliography{ICUAS-Ref}

\end{document}